\documentclass[11pt]{article}
\usepackage{amsmath,amssymb,epsfig,bm,slashbox}
\begin{document}

\date{\empty}

\title{\textbf{Slow decay of magnetic fields in open Friedmann universes}}

\author{John D. Barrow$^1$ and Christos G. Tsagas$^2$\\ {\small $^1$~DAMTP, Centre for Mathematical Sciences}\\ {\small University of Cambridge, Wilberforce Road, Cambridge CB3 0WA, UK}\\ {\small $^2$~Section of Astrophysics, Astronomy and Mechanics, Department of Physics}\\ {\small Aristotle University of Thessaloniki, Thessaloniki 54124, Greece}}

\maketitle

\begin{abstract}
We describe how magnetic fields in Friedmann universes can experience superadiabatic growth without departing from conventional electromagnetism. The reason is the relativistic coupling between vector fields and spacetime geometry, which slows down the decay of large-scale magnetic fields in open universes, compared to that seen in perfectly flat models. The result is a large relative gain in magnetic strength during the pre-galactic era that can lead to astrophysically interesting $B$-fields, even if our universe is only marginally open today.\newline\newline PACS numbers: 98.80.-k; 98.62.En; 98.65.Dx
\end{abstract}

Magnetic fields in Friedmann universes are widely believed to decay adiabatically regardless of the electrical properties of the cosmic medium. Consequently, large-scale $B$-fields are expected to dilute as $a^{-2}$, where $a$ is the cosmological expansion scale factor. This widespread perception has its roots in the conformal invariance of Maxwell's equations and the conformal flatness of the Friedmann-Robertson-Walker (FRW) spacetimes. The two are thought to guarantee that the rescaled magnetic vector $\mathcal{B}_{a}= a^{2}B_{a}$ evolves as in Minkowski space. This then ensures that the magnetic flux remains conserved and consequently that $B_{a}\propto a^{-2}$ irrespective of the electric properties of the universe. Following~\cite{W}, conformally flat spacetimes, like de Sitter space and the FRW models, can be written as time-dependent rescalings of Minkowski space. Then, the conformal triviality of Maxwell's theory guarantees that, when written on an FRW or a de Sitter background, the wave equation for the n-th Fourier mode of the rescaled magnetic vector $\mathcal{B}_a=a^2B_a$ takes the Minkowski-like form\footnote{To avoid confusion, we note that $B_a$ is the magnetic vector, $\mathcal{B}_a=a^2B_a$ is the rescaled magnetic vector, $B_{(\mathrm{n})}$ and $\mathcal{B}_{(\mathrm{n})}$ are their associated Fourier modes and $\rho_B=B^2/8\pi$ (with $B^2=B_aB^a$) is the magnetic energy density.}
\begin{equation}
\mathcal{B}_{(\mathrm{n})}^{\prime\prime}+ \mathrm{n}^2\mathcal{B}_{(\mathrm{n})}= 0\,,  \label{eq:W}
\end{equation}
with the primes indicating conformal-time derivatives. The above guarantees that $B_a\propto a^{-2}$ and therefore an adiabatic decay for the $B$-field. Hence the belief that to modify the $a^{-2}$-law and achieve a superadiabatic-type of magnetic amplification we need to abandon either the FRW models or conventional electromagnetism.\footnote{Superadiabatic amplification is a concept originally introduced in gravitational wave studies~\cite{G} and usually implies a magnetic decay-rate slower than the adiabatic one and not actual amplification.} The usual choice is to follow the latter route~\cite{TW}, and the literature contains a plethora of mechanisms that slow the adiabatic depletion of magnetic fields down by departing from standard electromagnetic theory (see~\cite{R} for a representative though incomplete list). This is not always necessary, however, because the argument leading to Eq.~(\ref{eq:W}) -- and the equation itself -- holds only when the FRW model has flat spatial sections (or when the space is exactly de Sitter).

All three Friedmann universes are conformally flat but they are not
identical. Differences in the geometry of their 3-spaces ensure that the conformal factor of the spatially curved models has an additional spatial dependence and therefore it no longer coincides with the cosmological scale factor. The associated line elements have the general form~\cite{S,IKM}
\begin{equation}
\mathrm{d}s^{2}= \alpha^{2}(\xi,R) \left[-\mathrm{d}\xi^{2}+\mathrm{d}R^{2} +R^{2}\mathrm{d}\Omega^{2}\right]\,,  \label{eq:cFRW}
\end{equation}
where $\alpha=\alpha(\xi,R)$ is the conformal factor and $\mathrm{d}\Omega^{2}=\mathrm{d}\theta^{2} +\sin^2\theta\mathrm{d}\phi^{2}$ (see~\cite{IKM} for details -- particularly on the open-FRW case). The above shows that Friedmann models with non-Euclidean spatial geometry cannot be written as simple, time-dependent rescalings of Minkowski space. For our purposes, this is the key difference between the flat and the rest of the FRW cosmologies. Putting it in geometrical terms, there is no global one-to-one correspondence between curved FRW models and Minkowski space: the conformal transformations mapping the associated spacetimes are only local~\cite{S}. Consequently, a Minkowski-like evolution for $\mathcal{B}_{a}$ is not a priori guaranteed in these models. Rescaling the magnetic field with a space-independent conformal factor does not work in the case of non-zero 3-curvature, as it does on a spatially flat FRW background, and the associated wave equation need not take the form of (\ref{eq:W}). Thus, the rescaled $\mathcal{B}$-field should show a Minkowski-like behaviour only locally (i.e.~on small scales). On large scales, where the curvature effects are important, one would in principle expect to see deviations from the standard $B_{a}\propto a^{-2}$-law.

This is indeed what happens. Consider, for example, the simple case of a source-free electromagnetic field on a general FRW background. Inflation is believed to generate such classical electromagnetic fields, by stretching the associated small-scale quantum fluctuations to super-horizon scales. In the absence of sources, the magnetic component of the Maxwell field propagates according to the linear wave equation~\cite{T1}
\begin{equation}
\ddot{B}_{a}-\mathrm{D}^{2}B_{a}= -5H\dot{B}_{a}- 4H^{2}B_{a}+ {\frac{1}{3}}\,(\rho+3p)B_{a}-\mathcal{R}_{ab}B^{b}\,.  \label{eq:ddotBa}
\end{equation}
Here, $B_{a}$ is the magnetic vector measured in a frame moving with 4-velocity $u_{a}$. The latter defines the comoving (fundamental) observers and it is normalised so that $u_{a}u^{a}=-1$. Also, the scalars $\rho$ and $p$ represent the matter density and pressure respectively, while $\mathcal{R}_{ab}$ is the background 3-Ricci tensor (see~\cite{BMT} for further details). Finally, overdots indicate proper-time derivatives, $\mathrm{D}^{2}= h^{ab}\nabla_{a}\nabla_{b}$ is the 3-D Laplacian (with $h_{ab}= g_{ab}+u_{a}u_{b}$ projecting orthogonal to $u_{a}$) and $H=\dot{a}/a$ is the Hubble parameter. Expression (\ref{eq:ddotBa}) can be obtained either by linearising the nonlinear equation (40) given in~\cite{T1} (see also Eq.~(45) in the same paper), or by simply recasting the first-order relation (28) of~\cite{MB}. In addition, the electromagnetic field vanishes in the unperturbed FRW background, which frees our analysis from gauge-related ambiguities. The key quantity for our purposes in Eq.~(\ref{eq:ddotBa}) is the magneto-curvature term, $\mathcal{R}_{ab}B^{b},$ which results from the non-commutativity of covariant derivatives in non-Euclidean spaces. It reflects the fact that vector sources, like the Maxwell field, `feel' the curvature of space through the Ricci identities -- in addition to Einstein's equations.

Introducing the rescaled $\mathcal{B}_{a}=a^{2}B_{a}$ field and using the conformal-time variable ($\eta $ with $\dot{\eta}=1/a>0$ and $~^{\prime}=\mathrm{d}/\mathrm{d}\eta $), the Fourier decomposition of (\ref{eq:ddotBa}) leads to
\begin{equation}
\mathcal{B}_{(\mathrm{n})}^{\prime\prime}+ \mathrm{n}^{2}\mathcal{B}_{(\mathrm{n})}= -2K\mathcal{B}_{(\mathrm{n})}\,,  \label{eq:cBwave}
\end{equation}
since $R_{ab}=(2K/a^{2})h_{ab}$ to zero order. In the above expression, which does not explicitly depend on the matter component, $K=0,\pm1$ is 3-curvature index of the background Friedmann model and $B_{a}= B_{(\mathrm{n})}\mathcal{Q}_{a}^{(\mathrm{n})}$ is the harmonically decomposed magnetic vector, with $\mathrm{D}_{a}B_{\mathrm{n}}=0= \dot{\mathcal{Q}}_{a}^{(\mathrm{n})}= \mathrm{D}^a\mathcal{Q}_a^{(\mathrm{n})}$ and $\mathrm{D}^{2}\mathcal{Q}_{a}^{(\mathrm{n})}= -(\mathrm{n}/a)^{2}\mathcal{Q}_{a}^{(\mathrm{n})}$. The Laplacian eigenvalue takes continuous values, with $\mathrm{n}^{2}\geq 0$, when $K=0,-1$ and discrete ones, with $\mathrm{n}^{2}\geq 3$, for $K=+1$. When $K=0$, the right-hand side of (\ref{eq:cBwave}) vanishes and we recover Eq.~(\ref{eq:W}). Otherwise, we need to account for the effects of the background geometry.

The curvature term on the right-hand side of expression (\ref{eq:cBwave}) can in principle modify the adiabatic decay-law of the $B$-field. For a $K=-1$ background, in particular,
\begin{equation}
\mathcal{B}_{(\mathrm{n})}^{\prime\prime}+ \left(\mathrm{n}^{2}-2\right)\mathcal{B}_{(\mathrm{n})}= 0\,,  \label{eq:-1cBwave}
\end{equation}
with $\mathrm{n}^{2}\geq0$. On large enough scales, with $\mathrm{n}^{2}<2$, the solutions to (5) no longer have the standard wave-like nature associated with the flat and the closed FRW hosts. These wavelengths include what one may regard as the largest subcurvature modes (i.e.~those with $1\leq\mathrm{n}^{2}<2$) and the supercurvature scales (having $0<\mathrm{n}^{2}<1$). Eigenvalues with $\mathrm{n}^{2}=1$ correspond to the curvature length with physical wavelength $\lambda=a$. Well inside this scale, the
3-space is practically flat, but beyond it the curvature dominates. Note that, although they are often omitted, supercurvature modes are necessary if we want perturbations with correlation lengths bigger than the curvature radius (see~\cite{LW} for further discussion). Here we will focus on the largest subcurvature modes. Let us introduce, for convenience, the scale-parameter $\mathrm{k}^{2}=2-\mathrm{n}^{2}$ with $0<\mathrm{k}^{2}<2$.
Then, the largest subcurvature scales correspond to the range $0<\mathrm{k}^{2}\leq1$, while the interval $1<\mathrm{k}^{2}<2$ contains the supercurvature lengths. In the new notation, the solution of Eq.~(\ref{eq:-1cBwave}) reads
\begin{equation}
B_{(\mathrm{k})}= a^{-2}\left[\mathcal{C}_{1}\cosh(|\mathrm{k}|\eta)+
\mathcal{C}_{2}\sinh(|\mathrm{k}|\eta)\right]= a^{-2}\,[\mathcal{C}_{3}\mathrm{e}^{|\mathrm{k}|\eta}+ \mathcal{C}_{4}\mathrm{e}^{-|\mathrm{k}|\eta}]\,,  \label{eq:-1B}
\end{equation}
where the $\mathcal{C}_{i}$s are the integration constants. As we will show next, magnetic fields obeying the above evolution law can experience superadiabatic amplification, without modifying conventional electromagnetism and despite the conformal flatness of their FRW host.

\begin{figure}[tbp]
\begin{center}
\includegraphics[height=5in,width=2in,angle=-90]{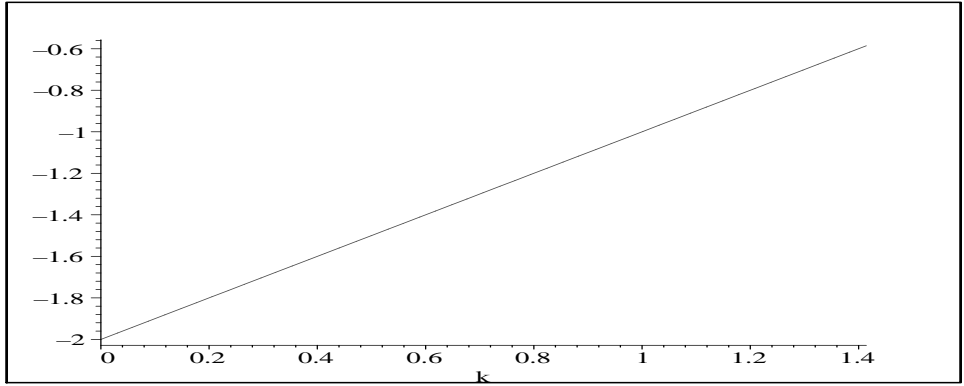}\quad
\end{center}
\caption{The ratio $\mathrm{d}(\ln B)/\mathrm{d}(\ln a)$ for the dominant magnetic mode (vertical axis) versus the scale parameter k (horizontal axis), according to solution (\protect\ref{eq:Milne}). Superadiabatic amplification occurs on all scales with $0<\mathrm{k}<\sqrt{2}$. The $\mathrm{k}=1$ value gives the curvature scale where $B\propto a^{-1}$. Stronger amplification occurs on supercurvature lengths, with $1<\mathrm{k}<\sqrt{2}$. Thus, as $\mathrm{k}\rightarrow\sqrt{2}$ and we approach infinite wavelengths, $B\propto a^{\sqrt{2}-2}$. At the $\mathrm{k}=0$ limit the $B$-field is well inside the curvature radius and the adiabatic decay is restored.}\label{fig:1}
\end{figure}

Suppose that the background model is a Milne-type universe: a vacuum, spatially open FRW spacetime with $a=t$. The latter immediately translates to $\mathrm{e}^{\eta}\propto a$, which substituted into solution (\ref{eq:-1B})
leads to
\begin{equation}
B_{(\mathrm{k})}= C_{1}a^{|\mathrm{k}|-2}+ C_{2}a^{-|\mathrm{k}|-2}\,,  \label{eq:Milne}
\end{equation}
with $C_{1,2}=$~constant. Consequently, all magnetic fields spanning lengths with $0<\mathrm{k}^{2}<2$ are superadiabatically amplified and their amplification strengthens with increasing scale. Close to the curvature-scale threshold, that is for $\mathrm{k}\rightarrow1$, the dominant mode is $B_{(1)}\propto a^{-1}$; a rate considerably slower than the adiabatic $a^{-2}$-law. The latter is only restored in the $\mathrm{k}=0$ limit, namely on small scales where the curvature effects are no longer important. Even stronger amplification is achieved on supercurvature scales, with $B_{(\mathrm{k})}\propto a^{\sqrt{2}-2}$ at the homogeneous limit (i.e.~as $\mathrm{k}\rightarrow\sqrt{2}$ -- see Fig.~\ref{fig:1} for a summary). The Milne universe is probably the simplest, but not the only FRW background, that supports magnetic amplification of the superadiabatic-type. Solution (\ref{eq:-1B}) leads to similar results in open universes with $p=\rho/3$ as well. To verify this recall that the scale factor of a radiation-dominated Friedmannian spacetime with $K=-1$ evolves according to $a\propto\sinh\eta$ (e.g.~see~\cite{B}). Then, focusing on the curvature scale for simplicity, Eq.~(\ref{eq:-1B}) ensures that a magnetic field with $\mathrm{k}=1$ never decays faster than $B_{(1)}\propto a^{-1}$ and is therefore superadiabatically amplified.\footnote{One can easily verify that, on the curvature scale (i.e.~for $\mathrm{n}^{2}=1$), a magnetic field evolving like $B_{a}\propto a^{-1}$ is a solution of Eq.~(\ref{eq:ddotBa}) when $\rho=0$ and also when $p=\rho/3$. The former equation of state corresponds to a Milne universe and the latter to a radiation-dominated FRW model (with $K=-1$ in our case).}

The Milne and the radiation-dominated Friedmann universe serve as very straightforward mathematical counterexamples, showing that the conformal flatness of the FRW spacetimes alone cannot guarantee an adiabatic decay for cosmological magnetic fields. For the purposes of physical cosmology, the Milne universe is the future attractor of (conventional) open FRW models. Thus, solution (\ref{eq:Milne}) could describe the late-time evolution of cosmological $B$-fields. The radiation example, on the other hand, refers to the early universe and can drastically increase the expected strength of large-scale magnetic fields. The high conductivity of the pre-recombination plasma, however, means that we cannot a priori ignore the electric currents in Eq.~(\ref{eq:ddotBa}). At the infinite conductivity limit, in particular, these currents are expected to dominate, eliminate the electric fields, and cause the magnetic component to freeze-in and dissipate adiabatically. Nevertheless, causality should confine the effects of the post-inflationary electric currents inside the horizon and therefore leave Eq.~(\ref{eq:ddotBa}) unaffected on super-Hubble scales. Recall that our electromagnetic field crossed the horizon before the onset of the radiation era, during the poorly conductive epoch of inflation.

To estimate the implications of the above described effects for primordial magnetic fields, it is important to note that large-scale $B$-fields experience an analogous superadiabatic type of amplification in open FRW universes with an inflationary (i.e.~$p=-\rho$) equation of state~\cite{TK}. In these models, fields with coherence lengths close -- and beyond -- the curvature scale also experience superadiabatic amplification triggered by the same 3-curvature effects we described above. To be precise, $B$-fields near the curvature scale were found to obey the evolution law
\begin{equation}
B_{(1)}= C_{3}\left(1-\mathrm{e}^{2\eta}\right)a^{-1}+ C_{4}\mathrm{e}^{-\eta }a^{-2}\,,  \label{eq:-1Bf}
\end{equation}
where $\eta<0$ and $C_{3,4}$ constants (see~\cite{TK} for details). Thus, for most of the inflationary phase (i.e.~as long as $\eta\ll0$), we have $B_{(1)}\propto a^{-1}$ and the field is superadiabatically amplified. The adiabatic decay rate is recovered only at the end of inflation, as $\eta\rightarrow0^{-}$. We emphasise that the reduction in the magnetic decay-rate is possible because inflation in curved Friedmann models does not lead to a globally flat de Sitter space. Although inflation can dramatically increase the curvature radius of the universe, it does not change its spatial geometry. Unless the universe was perfectly flat from the beginning, there will always be a scale where the 3-curvature effects are important.\footnote{In probabilistic terms, an exactly flat universe is a set of measure zero and therefore physically unrealisable.} It is on these lengths that the $B$-fields can be superadiabatically amplified.

Following~\cite{TK}, a magnetic field that survived a de Sitter-type inflation in a FRW cosmology with $K=-1$ has energy density $\rho_{B}\sim10^{-51}\left(M/10^{17}\mathrm{GeV}\right)^{8/3} \left(T_{RH}/10^{9}\mathrm{GeV}\right)^{-2/3}\lambda_{Mpc}^{-2}\rho_{\gamma}$, and a current typical (comoving) strength from $\sim10^{-35}$ to $\sim10^{-33}$ Gauss, depending on the parameters of the defining inflationary model. Note that $M$ is the energy scale of inflation, $T_{RH}$ is the reheating temperature, $\rho_{\gamma}$ is the
radiation energy density and $\lambda_{Mpc}$ is the current scale of the amplified $B$-field.\footnote{Assuming $M\sim10^{17}$~GeV and $H_{0}\simeq70$~km/sec\thinspace Mpc, one obtains a residual $B$-field of the order $10^{-35}$~G when $T_{RH}\sim10^{9}$~GeV. A reheating temperature close to $10^{3}$~GeV, on the other hand, leads to $B\sim10^{-33}$~G. In general, the higher the scale of inflation and the lower the reheating temperature the stronger the amplification~\cite{TK}.} The latter is close to the curvature scale which, for a marginally open universe -- with $1-\Omega\sim10^{-2}$ today, lies between $10^{4}$ and $10^{5}$~Mpc. These scales are far larger than the minimum magnetic length required for the galactic dynamo to work.\footnote{The dynamo mechanism requires magnetic seeds with a (collapsed) coherence length close to 100~pc. This corresponds to a comoving (pre-collapse) scale of approximately 10~Kpc.} Nevertheless, once galaxy formation starts, the fluid motion should force the magnetic force lines to break up and reconnect on lengths similar to the size of a collapsing protogalaxy. Magnetic fields in the range of $10^{-35}$ to $10^{-33}$ Gauss are far stronger than any other conventional $B$-field that went through an epoch of inflation. So far, similar strengths have only been achieved outside standard electromagnetism. Moreover, seeds around $10^{-34}$~G (or less) can sustain the galactic dynamo if our Universe is currently dark-energy dominated~\cite{DLT}.

The above quoted strengths assume that the magnetic component freezes-in after the end of inflation and that the ratio $\rho_B/\rho_{\gamma}$ remains constant throughout the radiation epoch and until today~\cite{TK}. Hence, the residual $B$-field will increase further if it is superadiabatically amplified during the radiation era as well. Following our earlier discussion, this is possible for superhorizon-sized magnetic fields of inflationary origin, because they are not affected by the electric currents of the post-inflation universe. In such a case, an evolution law of $a^{-1}$ during the whole of the radiation epoch, will add several orders of magnitude to the residual $B$-field. For example, the above quoted magnetic seed of $10^{-35}$~G will `grow' further during the radiation era. This field, which has length close to the curvature radius and corresponds to $T_{RH}\sim10^{9}$~GeV, will be superadiabatically amplified by $T_{RH}/T_{rec}\sim10^{19}$ orders of magnitude and reach a comoving strength of up to $10^{-16}$~G. Note that, when $1-\Omega_{0}\sim10^{-2}$, magnetic fields spanning lengths near the curvature scale of the universe, remain outside the horizon (and therefore are superadiabatically amplified) throughout the radiation era. Also, although they leave primordial nucleosynthesis and the Cosmic Microwave Background unaffected~\cite{GR}, magnetic seeds of the order of $10^{-16}$~G are astrophysically important because they can sustain the galactic dynamo even in conventional universes with zero dark energy~\cite{ZRS}. With this in mind, a more detailed (most likely numerical) analysis is necessary to establish the full spectrum of the residual $B$-field. Here we have confined ourselves to effects close to the curvature scale, which probably means that the estimated magnetic strengths are the maximum possible.

Large-scale magnetic fields have been observed almost everywhere in the universe. From galaxies and galaxy clusters, to super-clusters and remote high-redshift protogalactic structures, observations have repeatedly detected coherent $B$-fields of micro-Gauss strength. Despite this widespread presence, however, the origin of cosmic magnetism is still unknown and a matter of debate. Inflation has long been seen as our best mechanism for generating large-scale magnetic fields, because it naturally leads to superhorizon-sized correlations. The resulting fields, however, were always considered too weak to have any astrophysical significance. This was attributed to the conformal invariance of Maxwell's theory and to the conformal flatness of the FRW spacetimes. The two were believed to guarantee an adiabatic decay for any cosmological $B$-field, irrespective of the electric properties of the universe.

Contrary to this widespread belief, however, magnetic fields on FRW backgrounds do not always experience an adiabatic, Minkowski-like depletion. Departures from the $a^{-2}$-law, without breaking away from conventional electromagnetism, are possible in FRW models with non-Euclidean spatial hypersurfaces, because those spacetimes are not globally conformal to Minkowski space. This geometrical subtlety, which has long been known within the relativity community, but has been largely overlooked in studies of cosmological magnetic fields, is central to our report. Together with the vector nature of the Maxwell field, which brings the 3-curvature into play, it is essential in understanding the behaviour of large-scale magnetic fields in FRW cosmologies and avoids the need to deviate from Maxwellian electromagnetism in our quest for astrophysically relevant cosmological magnetic seeds.

\end{document}